\documentclass[aps,pra,reprint,groupedaddress,showpacs,floatfix,longbibliography]{revtex4-2}
\usepackage{graphicx}
\usepackage{amssymb,amsmath}
\usepackage{dcolumn}
\usepackage{bm}
\usepackage{upgreek}
\usepackage[colorlinks=true, linkcolor=blue, citecolor=blue, urlcolor=blue]{hyperref}
\usepackage[mathlines]{lineno}
\usepackage{multirow}
\usepackage{appendix}
\usepackage{color}
\begin{document}
	\title{From flat to narrow bands: Engineering quantum emission in a one-dimensional Lieb lattice}
	\author{Zhiyong Liu}
	\affiliation{State Key Laboratory of Quantum Optics and Quantum Optics Devices, Institute of Laser Spectroscopy, Shanxi University, Taiyuan 030006, China \\ and Collaborative Innovation Center of Extreme Optics, Shanxi University, Taiyuan 030006, China}
	\author{Yue Sun}
	\affiliation{State Key Laboratory of Quantum Optics and Quantum Optics Devices, Institute of Laser Spectroscopy, Shanxi University, Taiyuan 030006, China \\ and Collaborative Innovation Center of Extreme Optics, Shanxi University, Taiyuan 030006, China}
	\author{Ying Hu}
	\email{huying@sxu.edu.cn}
	\affiliation{State Key Laboratory of Quantum Optics and Quantum Optics Devices, Institute of Laser Spectroscopy, Shanxi University, Taiyuan 030006, China \\ and Collaborative Innovation Center of Extreme Optics, Shanxi University, Taiyuan 030006, China}
	
	\begin{abstract}
		We develop a comprehensive theoretical framework that unifies quantum emission dynamics in one-dimensional Lieb lattices, bridging the gap between ideal flat-band coherence and realistic narrow-band dissipation. By coupling an emitter to sublattices with finite flat-band wavefunction overlap, we activate a collective, size-independent interaction fundamentally distinct from dispersive-band processes. Controllably breaking lattice symmetry transforms the flat band into a narrow dispersive band, enabling a continuous crossover from non-Markovian to Markovian dynamics governed by the competition between coupling strength and engineered bandwidth. Crucially, we derive explicit scaling laws that provide a quantitative blueprint for tuning spontaneous emission from coherent trapping to Markovian decay. Our work provides a unified framework that connects idealized flat-band physics to emerging narrow-band platforms such as moir$\rm\acute{e}$ photonic crystals, offering a practical toolkit for interpreting experiments and engineering quantum emission in structured photonic environments.
	\end{abstract}
	
	\maketitle
	\section{Introduction}
	Photonic crystals, with their engineered band gaps and density of states (DOS), provide a powerful platform for controlling light-matter interactions~\cite{gonzalez2024light,rivera2020light}. Such control enables a wide range of phenomena, including inhibited~\cite{noda2007spontaneous,asenjo2017exponential,lodahl2015interfacing,kopp1998low} and enhanced spontaneous emission~\cite{pelton2015modified,lodahl2004controlling,fan1997high,lund2008experimental}, the formation of atom-photon bound states~\cite{sheremet2023waveguide,weimann2017topologically,shi2016bound}, and the simulation of quantum many-body physics~\cite{hartmann2006strongly,douglas2015quantum,gonzalez2015subwavelength}. This ability to tailor the photonic dispersion relation is central to these effects, particularly in low-dimensional nanostructures such as photonic crystal waveguides~\cite{frandsen2006photonic,young2015polarization} and metasurfaces~\cite{yu2014flat,solntsev2021metasurfaces,luk2010fano}. This precise engineering has opened new avenues for quantum information processing and the exploration of novel quantum phases~\cite{hennessy2007quantum,lodahl2015interfacing,atature2018material}.
	
	Among structured photonic environments, systems with flat bands, characterized by dispersionless energy eigenvalues and a divergent DOS, represent a particularly powerful paradigm~\cite{leykam2018artificial,ren2025far,zong2016observation,tang2020photonic,mukherjee2018experimental,mukherjee2015observation,flach2014detangling}. Early seminal works laid the foundation for understanding flat-band localization and its implications for phenomena like disorder-induced localization~\cite{leykam2017localization,zeng2024transition} and the nontrivial topological phase~\cite{pal2018nontrivial}. Subsequent advances in photonic systems, including the Lieb lattice~\cite{mukherjee2015observation,vicencio2015observation,flach2014detangling,PhysRevA.87.023614,tevcer2025flat}, diamond-chain~\cite{marques2024impurity} and Kagome lattice~\cite{danieli2024flat,zong2016observation,poblete2021photonic,zhang2019kagome}, have vividly demonstrated the observation and engineering of flat bands for tailoring light-matter interactions. Recent studies revealed exotic nonlinear dynamics from flat-band modes~\cite{di2019nonlinear}, while work in coupled waveguide arrays and metasurfaces demonstrated robust flat-band transport and enhanced light-matter coupling~\cite{eyvazi2025flat,vicencio2025observation,di2025dipole}.
	
	In realistic experimental systems, ideally flat bands are difficult to realize. Recent breakthroughs in engineered photonic structures, particularly moir$\rm\acute{e}$ photonic crystals, successfully created bands with extremely narrow dispersion, opening a new frontier for controlling light-matter interactions~\cite{yang2023photonic,wang2025moire,dong2021flat}. The bandwidths of the narrow bands in these systems are much smaller than the band gaps between the narrow bands and other bands. Experiments coupling a single quantum emitter to such narrow bands demonstrated remarkable phenomena, including dramatically accelerated spontaneous emission~\cite{wang2025moire}, which seemed unexpected from flat-band physics. In this scenario, the emitter no longer interacts with a set of perfectly dispersionless states, as in an ideal flat band, but rather with a continuum of modes that, while narrow, possess a finite bandwidth. This distinction, though subtle, fundamentally alters the nature of light-matter coupling, which we systematically explore in this work.
	
	\begin{figure}[t]
		\includegraphics[width=1.0\columnwidth]{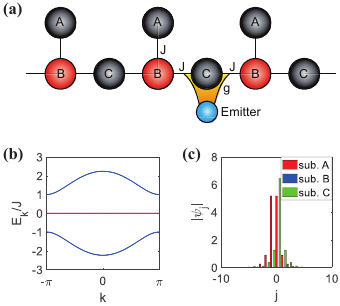}
		\caption{(Color online) Schematic, band structure, and flat-band wavefunctions of the 1D Lieb lattice. (a) Schematic illustration of a quantum emitter (blue sphere) locally coupled to the $C$ sublattice of the 1D Lieb lattice model. (b) The corresponding band structure, featuring a central flat band (red line) at energy $E_k=0$ and two dispersive bands (blue lines). (c) The spatial profile of the flat-band wavefunction in real space $|\psi_j|$ for a lattice with $N=120$ unit cells. The red, blue, and green bars represent the wavefunction amplitude on the $A$, $B$, and $C$ sublattices, respectively, highlighting the compact localization and the dark-state nature of the $B$ sublattice.}
		\label{fig1}
	\end{figure}

	In this work, we establish precisely such a framework by systematically exploring quantum emission in a tunable one-dimensional Lieb lattice. Our approach is designed to deconstruct, in a controlled manner, the distinct physical mechanisms that govern emitter dynamics in flat-band versus narrow-band environments. By exploiting the sublattice degree of freedom, we selectively activate or suppress coupling to the flat band, thereby isolating its collective, size-independent interaction from conventional dispersive-band processes. Crucially, we go beyond the pristine flat-band limit by controllably introducing symmetry-breaking perturbations, enabling a continuous transition from a perfectly dispersionless band to a narrow-band regime that mirrors realistic platforms such as moir$\rm\acute{e}$ photonic crystals. The central thrust of our work is to identify and quantitatively map the competition between the intrinsic emitter-bath coupling strength and the engineered bandwidth, the key parameters that dictate the emitter's dynamics. Through the derivation and validation of explicit scaling laws, our framework provides not merely a description, but a predictive toolkit for tailoring emission dynamics across regimes. This work thereby bridges the conceptual divide between idealized flat-band physics and the realities of experimental narrow-band systems, offering both a unifying physical picture and practical design principles for engineering quantum light-matter interactions in structured photonic environments.
	
	\section{Emitter couples to one-dimensional Lieb lattice with pristine flat band}
	In this section, we investigate the spontaneous emission dynamics of a quantum emitter locally coupled to the one-dimensional (1D) Lieb lattice shown in Fig.~\ref{fig1}(a). We begin by presenting the model. The total Hamiltonian is given by $H=H_s+H_b+H_c$, describing the emitter, the bath, and their interaction, respectively. In the rotating frame with respect to the central frequency of the bath modes, the emitter Hamiltonian is
	\begin{equation}
		H_s=\Delta a^\dagger a,
		\label{eq1}
	\end{equation}
	where $\Delta$ denotes the emitter detuning. The photonic bath is a 1D Lieb lattice described by the tight-binding Hamiltonian
	\begin{equation}
		H_b=\sum_j J\left(b_j^{A\dagger}b_j^B+b_j^{B\dagger}b_j^C+b_j^{B\dagger}b_{j+1}^C+\text{h.c.}\right),
		\label{eq2}
	\end{equation}
	where $J$ is the nearest-neighbor hopping amplitude, and $b_j^\gamma$ ($b_j^{\gamma\dagger}$) is the annihilation (creation) operator for site $\gamma(=A,B,C)$ in the $j$th unit cell. Diagonalizing $H_b$ yields a band structure consisting of three bands: a central flat band at energy $E_k=0$, flanked by two dispersive bands with energies $E_k=\pm J\sqrt{3+2\cos(k)}$, as shown in Fig.~\ref{fig1}(b).
	
	The flat-band wavefunction in real space can be expressed in Bloch form as
	\begin{equation}
		\psi_j=\frac{1}{\sqrt{N}}\sum_k\frac{e^{ikj}}{\sqrt{3+2\cos(k)}}\left(\begin{array}{c}
			-1-e^{ik} \\ 0 \\ 1
	 	\end{array}\right),
	 	\label{eq3}
	\end{equation}
 	where $j$ indexes the unit cells. The wavefunction plotted in Fig.~\ref{fig1}(c) is computed directly from Eq.~(\ref{eq3}) for a lattice with $N=120$ unit cells, which shows that the flat band supports localized states that are confined to few unit cells in real space. The key feature is that the flat-band wavefunction has exactly zero amplitude on the $B$ sublattice, with finite amplitude only on the $A$ and $C$ sublattices. This fundamental distinction implies that when the emitter couples to the $A/C$ sublattice, it can interact with the flat-band modes, whereas when the emitter couples to the $B$ sublattice, it cannot couple to the flat-band modes. We now examine the spontaneous emission dynamics for these two distinct coupling scenarios in detail.
 	
 	\subsection{Emitter couples to $B$ sublattice}
 	To establish a baseline for understanding the unique role of flat bands, we first consider a scenario where the emitter is coupled to the $B$ sublattice at the $j=0$ unit cell. The interaction Hamiltonian is
 	\begin{equation}
 		H_c=g\left(a^\dagger b_0^B+b_0^{B\dagger}a\right),
 		\label{eq4}
 	\end{equation}
 	where $g$ is the emitter-bath coupling strength. Since the $B$ sublattice hosts the dark state of the flat band [see Eq.~(\ref{eq3})], the emitter does not directly couple to the flat-band modes.
 	
 	The spontaneous emission dynamics is characterized by the temporal evolution of the emitter's excited-state population, $P(t)=\langle a^\dagger(t)a(t)\rangle$ with $P(0)=1$, which can be written as
 	\begin{equation}
 		P(t)=\left|G_s(t)\right|^2.
 		\label{eq5}
 	\end{equation}
 	This population can be expressed as the square of the modulus of the retarded Green's function, $G_s(t)=-i\langle0|a(t)a^\dagger(0)|0\rangle$ for $t>0$, where the expectation value is taken over the vacuum state. To analyze this dynamics, we employ the Lehmann spectral representation, which provides a powerful framework for decomposing the Green's function into contributions from bound states and the continuum. In this representation, $G_s(t)$ is given by
 	\begin{equation}
 		\begin{split}
 			G_s(t)&=D_s(t)+D_b(t) \\
 			&=\sum_sZ_se^{-iE_st}+\int_{\textrm{bath}}\textrm{d}\omega A(\omega)e^{-i\omega t},
 		\end{split}
 		\label{eq6}
 	\end{equation}
	where $D_s(t)$ and $D_b(t)$ correspond to the contributions from the discrete poles (bound states) and the branch cut (continuum bath) of the Green's function in the frequency domain $G_s(\omega)$, respectively. The single-particle Green's function in $G_s(\omega)$ can be expressed as
	\begin{equation}
		G_s(\omega)=\frac{1}{\omega-\Delta-\Sigma_s(\omega)},
		\label{eq7}
	\end{equation}
	where the self-energy $\Sigma_s(\omega)$ captures the influence of the bath. For the emitter coupled to the $B$ sublattice, the self-energy is found to be (derived in Appendix~\ref{appendix:A})
	\begin{equation}
		\Sigma_s(\omega)=\frac{g^2\omega\theta(1-|z_-|)}{\sqrt{(\omega^2-3J^2)^2-4J^4}}-\frac{g^2\omega\theta(1-|z_+|)}{\sqrt{(\omega^2-3J^2)^2-4J^4}},
		\label{eq8}
	\end{equation}
	where $\theta(1-|z_\pm|)$ is the step function with $z_\pm=\frac{\omega^2-3J^2\pm\sqrt{(\omega^2-3J^2)^2-4J^4}}{2J^2}$. The residue associated with a bound state at energy $E_s$ is $Z_s=\sum_s\left(1-\partial_\omega\Sigma_s(\omega)|_{\omega=E_s}\right)^{-1}$ and the spectrum function of the single-particle Green's function is $A(\omega)=\frac{1}{\uppi}\textrm{Im}\left[G_s(\omega+i0^+)\right]$. With $Z_s$ and $A(\omega)$ determined as above, the full spontaneous emission dynamics can be evaluated via Eq.~(\ref{eq6}).
	
	\begin{figure}[t]
		\includegraphics[width=1.0\columnwidth]{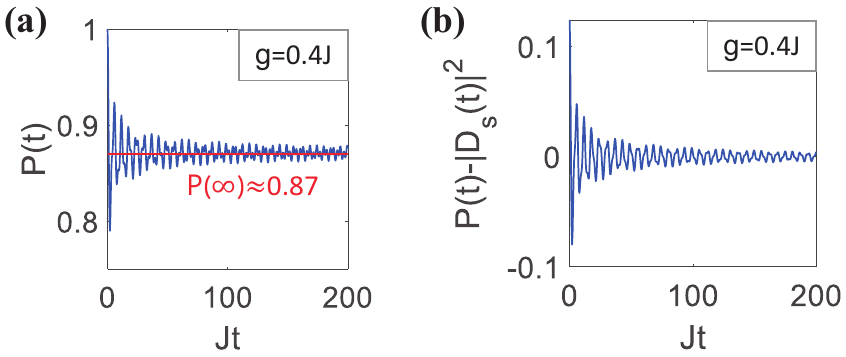}
		\caption{(Color online) Spontaneous emission dynamics for an emitter coupled to the $B$ sublattice with $\Delta=0$ and $g=0.4J$. (a) Emitter population $P(t)$ as a function of time. The population saturates to a steady-state value $P(t)|_{t\to\infty}\approx0.87$, demonstrating strong population trapping. (b) The oscillatory component $P(t)-\left|D_s(t)\right|^2$, quantifying the interference between bound states and the continuum.}
		\label{fig2}
	\end{figure}
	
	The formation of bound states is confirmed by solving $G_s^{-1}(E_s)=0$. For $\Delta=0$, the emitter resonates with the flat band, and we identify three bound states at energies $E_s=0$ and $E_s=\pm\sqrt{3J^2+\sqrt{4J^4+g^4}}$. For a coupling strength of $g=0.4J$, our analytical framework predicts a long-time population of $P(t)|_{t\to\infty}\approx0.87$. To verify this, we perform a direct numerical simulation of the spontaneous emission dynamics, $P(t)=\left|e^{-iHt}|\psi(0)\rangle\right|^2$, with the initial state $|\psi(0)\rangle$ representing the emitter in its excited state and the bath in the vacuum state. The excellent agreement between the analytical prediction and the numerical result is shown by the red line in Fig.~\ref{fig2}(a). The oscillations shown in Fig.~\ref{fig2}(b) stem from the interference between the contributions from the bound states and the continuum, quantified by $P(t)-\left|D_s(t)\right|^2$.
	
	This behavior can be further elucidated in the band representation. Here, the bath Hamiltonian is diagonalized as
	\begin{equation}
		H_b=\sum_{k,n}E_k^{(n)}\psi_k^{(n)\dagger}\psi_k^{(n)},
		\label{eq9}
	\end{equation}
	where $n$ is the band index and $E_k^{(\pm)}=\pm J\sqrt{3+2\cos(k)}$ is the energy dispersion of the dispersive band, and $\psi_k^{(\pm)}$ are the corresponding eigenmode annihilation operators. The interaction Hamiltonian becomes (derived in Appendix~\ref{appendix:B})
	\begin{equation}
		H_c=\sum_{k,n}g_k^{(n)}\left(a^\dagger\psi_k^{(n)}+\psi_k^{(n)\dagger}a\right),
		\label{eq10}
	\end{equation}
	where $g_k^{(n)}$ is the coupling strength between the emitter and the bath mode $(k,n)$. For the emitter coupled to the $B$ sublattice considered here, we find $g_k^{(\pm)}=\frac{g}{\sqrt{N}}\times\frac{\sqrt{1+\cos(k)}}{1+e^{-ik}}$ and crucially $g_k^{(0)}=0$, confirming that the flat band remains completely decoupled. The observed bound states and population trapping therefore originate solely from interaction with the dispersive bands, establishing a clean baseline for comparison with the case where flat-band coupling is present.
	
	\subsection{Emitter couples to $C$ sublattice}
	
	\begin{figure*}[t]
		\includegraphics[width=2.0\columnwidth]{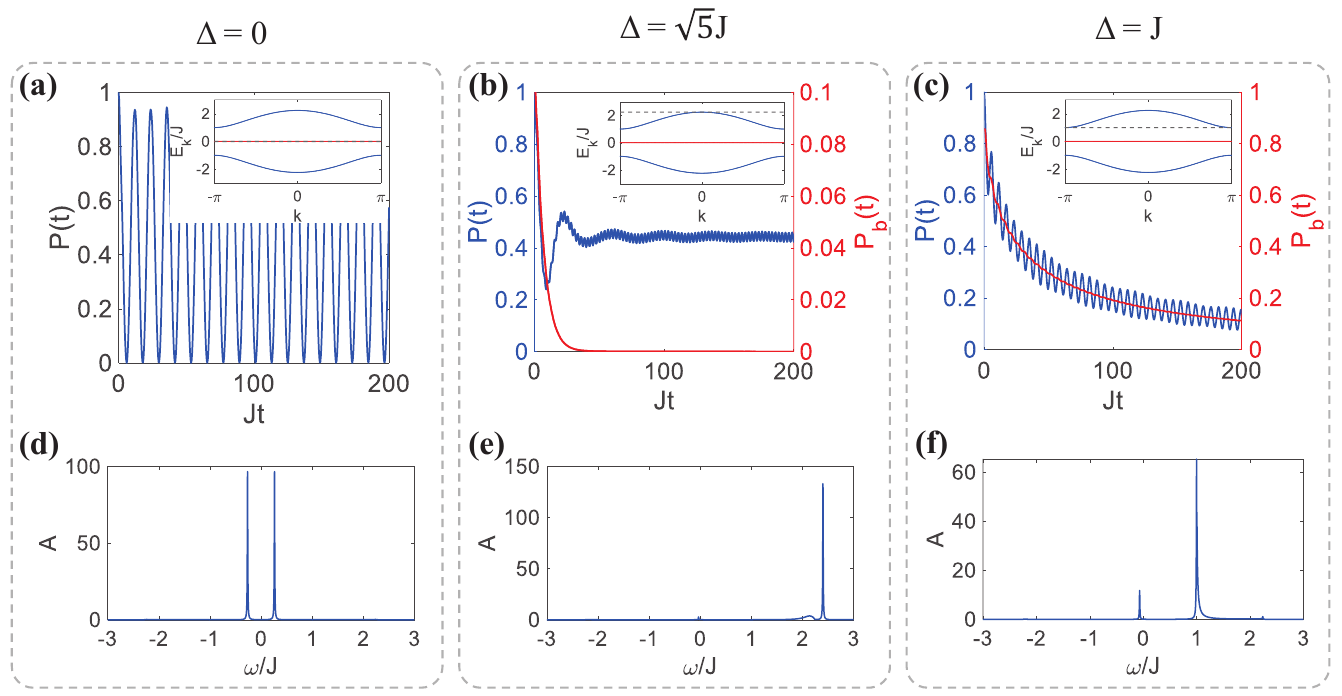}
		\caption{(Color online) Dynamics and spectral properties for an emitter coupled to the C sublattice with $g=0.4J$. (a)-(c) Spontaneous emission dynamics $P(t)$ (blue lines) for (a) $\Delta=0$, (b) $\Delta=\sqrt{5}J$ and (c) $\Delta=J$. The red lines represent the continuum contribution $P_b(t)=\left|D_b(t)\right|^2$. The dynamics exhibit a transition from (a) strong coherent trapping to (c) nearly complete decay. Insets: Schematic of the emitter detuning (gray dashed line) relative to the bath bands. (d)-(f) The corresponding spectral function $A(\omega)$ for each detuning case in (a)-(c). Sharp peaks outside the continuum correspond to bound states, while broad features originate from the dispersive band continuum.}
		\label{fig3}
	\end{figure*}

	Having established the baseline dynamics mediated solely by dispersive bands, we now investigate the case where the emitter couples to the $C$ sublattice at the $j=0$ unit cell. This configuration allows us to probe how the presence of the flat band, through its finite wavefunction overlap, fundamentally modifies the spontaneous emission dynamics. The intraction Hamiltonian is
	\begin{equation}
		H_c=g\left(a^\dagger b_0^C+b_0^{C\dagger}a\right),
		\label{eq11}
	\end{equation}
	where $g$ is the emitter-bath coupling strength. For this configuration, the self-energy is found to be (derived in Appendix~\ref{appendix:A})
	\begin{equation}
		\Sigma_s(\omega)=\frac{g^2(\omega^2-J^2)\theta(1-|z_-|)}{\omega\sqrt{(\omega^2-3J^2)^2-4J^4}}-\frac{g^2(\omega^2-J^2)\theta(1-|z_+|)}{\omega\sqrt{(\omega^2-3J^2)^2-4J^4}},
		\label{eq12}
	\end{equation}
	where $\theta(1-|z_\pm|)$ is the step function with $z_\pm=\frac{\omega^2-3J^2\pm\sqrt{(\omega^2-3J^2)^2-4J^4}}{2J^2}$. The bound states $E_s$ are again determined by $G_s^{-1}(E_s)=0$.
	
	For $\Delta=0$, the emitter resonates with the flat band, and two in-gap bound states emerge at energies $E_\pm\approx\pm0.264J$. The total residue of these bound states is $\sum_{s=\pm}Z_s\approx0.97$, indicating that the spontaneous emission dynamics is dominated by their contribution, $P(t)=\left|\sum_{s=\pm}Z_se^{-iE_s t}\right|^2=\left|2Z_s\sin\left(\left|E_s\right|t\right)\right|^2$. The coherent oscillations evident in the numerical results of Fig.~\ref{fig3}(a) originate from the two bound states. The oscillation period, $T=\frac{\uppi}{J\left|E_s\right|}\approx11.9/J$, determined by the bound-state energy, shows excellent agreement with the numerical simulation. Furthermore, the constant amplitude of these oscillations over a long time is consistent with the large total residue, signifying minimal population decay into the continuum. The spectral function $A(\omega)$, plotted in Fig.~\ref{fig3}(d), shows sharp peaks at the bound-state energies, while the flat band contributes negligible spectral weight, confirming its role in mediating the formation of bound states rather than irreversible decay.
	
	The formation mechanism of these bound states is fundamentally distinct from the $B$ sublattice case. Here, the flat band does not act as a passive reservoir but as a single, collective degree of freedom (derived in Appendix~\ref{appendix:B}). In band representation, the flat-band Hamiltonian becomes Eq.~(\ref{eq9}) with $E_k=0$. The interaction Hamiltonian in the diagonal basis is Eq.~(\ref{eq10}), where $g_k$ is the coupling constant between the emitter and the bath mode $k$. For the flat band, the coupling strength is $g_k^{(\textrm{flat})}=\frac{g}{\sqrt{N}}\times\frac{1}{\sqrt{3+2\cos(k)}}$, which can be collectively enhanced. In collective representation, the interaction Hamiltonian can be rewritten as $H_c=g_B\left(a^\dagger B+B^\dagger a\right)$, where $B=\frac{\sum_k g_k^{(\textrm{flat})}\psi_k}{\sqrt{\sum_k|g_k^{(\textrm{flat})}|^2}}$ is a collective mode encompassing the entire flat band, and the effective coupling strength $g_B=\frac{g}{5^{1/4}}$ is independent of system size. This collective enhancement leads to the steady coherent oscillations in dynamics [shown in Fig.~\ref{fig3}(a)]. Therefore, at $\Delta=0$, the spontaneous emission dynamics differs markedly depending on the coupling sublattice: it is dominated by bound states arising from the dispersive bands when the emitter couples to the $B$ sublattice, in contrast to the case of $C$ sublattice coupling, where the dynamics is governed by bound states formed via the flat band.
	
	We further explore the detuning dependence of the emitter's dynamics. While the DOS diverges at both the dispersive band edges and the flat band, the spontaneous emission dynamics exhibits markedly distinct characteristics. For $\Delta=\sqrt{5}J$, the emitter resonates with the upper edge of the dispersive band [inset of Fig.~\ref{fig3}(b)]. The dynamics calculated numerically, shown by the blue line in Fig.~\ref{fig3}(b), is consistent with the analytical result from Eq.~(\ref{eq6}). In this case, one bound state resides inside the gap ($E_-\approx-0.032J$) and another lies above the continuum ($E_+\approx2.402J$). The long-time dynamics is dominated by the two bound states, $\left|\sum_{s=\pm}Z_se^{-iE_s t}\right|^2$. The oscillation period is $T=\frac{2\uppi}{J\left|E_+-E_-\right|}\approx2.58/J$ and the amplitude of these oscillations is governed by the residue $Z_s\approx0.67$ of the bound state at $E_+\approx2.402J$, closer to the upper band edge, which are consistent with the numerical results [Fig.~\ref{fig3}(b)].
	
	An analogous discussion applies to the case of $\Delta=J$ [inset of Fig.~\ref{fig3}(c)], where one bound state lies below the flat band ($E_-\approx-0.067J$) and another above the upper dispersive band ($E_+\approx2.239J$). The dynamics calculated numerically, shown by the blue line in Fig.~\ref{fig3}(c), is consistent with the analytical result from Eq.~(\ref{eq6}). The oscillation period $T=\frac{2\uppi}{J\left|\Delta-E_-\right|}\approx5.9/J$ results from the interference between the bound state and the continuum. This analytical prediction is consistent with the numerical results shown in Fig.~\ref{fig3}(c). The amplitude of the long-time population governed by the residue $Z_-\approx0.059$ of the bound state $E_-\approx-0.067J$, closer to the lower band edge, indicating that nearly all photons in the emitter eventually radiate into the bath. From Figs.~\ref{fig3}(a) to \ref{fig3}(c), we observe that the emitter's spontaneous emission dynamics is highly sensitive to its detuning, enabling a transition from strong non-Markovian oscillations to nearly Markovian decay.
	
	The red lines in Figs.~\ref{fig3}(b) and \ref{fig3}(c) represent the continuum contribution $P_b(t)=\left|D_b(t)\right|^2$ by the second term of Eq.~(\ref{eq6}), which captures the spontaneous emission dynamics evolving from the initial condition. The decay rate of $P_b(t)$ is significantly larger for $\Delta=\sqrt{5}J$ than for $\Delta=J$. This difference can be understood by examining the momentum dependence of the coupling $g_k$ given in Eq.~(\ref{eq10}) together with the computed spectral features. More intuitively, the origin lies in the mirror symmetry of the bulk modes [calculated by Eq.~(\ref{eq3})]: the mode at $E=\sqrt{5}J$ (corresponding to $k=0$) is mirror-symmetric about the emitter site on the C sublattice, leading to a finite coupling strength $g_k$. In contrast, the mode at $E=J$ (corresponding to $k=\uppi$) is mirror-antisymmetric, forcing the wavefunction amplitude at the emitter site to vanish and thus strongly suppressing the coupling. For the dispersive band, the coupling strength is $g_k^{(\textrm{disp})}=\frac{g}{\sqrt{N}}\times\sqrt{\frac{1+\cos(k)}{3+2\cos(k)}}$, which is highly asymmetric and decreases monotonically with $k$ from 0 to $\uppi$ (derived in Appendix~\ref{appendix:B}). The spectral function $A(\omega)$ reveals that for $\Delta=\sqrt{5}J$ [shown in Fig.~\ref{fig3}(e)], two sharp peaks at the bound state energies coexist with a broad continuum near the upper band edge of the dispersive band ($k\approx0$), where the coupling strength $g_k^{(\textrm{disp})}$ is relatively large. In contrast, for $\Delta=J$ [shown in Fig.~\ref{fig3}(f)], the spectral weight is concentrated near the lower band edge ($k\approx\uppi$), where the coupling strength $g_k$ is small. Therefore, the decay rate of $P_b(t)$ is significantly larger for $\Delta=\sqrt{5}J$ than for $\Delta=J$.
	
	The rich detuning dependence revealed in Figs.~\ref{fig3}(a) to \ref{fig3}(c) finds a unified explanation in the contrasting nature of emitter-band coupling. Our band-representation analysis yields a key insight: The interaction with the flat band is mediated by a single, collective mode $B$, leading to a size-independent effective coupling strength $g_B$. In contrast, coupling to the dispersive bands involves a sum over extended Bloch modes, resulting in a size-dependent coupling. The coupling strength of the flat band $g_B$ is much larger than the dispersive band $g_k\propto\frac{1}{\sqrt{N}}$. Furthermore, both flat bands and the edges of dispersive bands possess a divergent DOS, yet they can give rise to completely different spontaneous emission dynamics for the emitter, owing to distinct coupling strengths. Thus, the detuning $\Delta$ acts as a continuous dial that reweights the emitter's interaction between the robust, collective flat-band channel and the dissipative, extended continua, offering a powerful mechanism to control the quantum dynamical regime.
	
	For completeness, the dynamics for an emitter coupled to the $A$ sublattice are presented in Appendix~\ref{appendix:C}, which exhibits properties similar to those of the $C$ sublattice case discussed above.
	
	\section{Emitter couples to modified 1D Lieb lattice with narrow band}
\begin{figure*}[t]
		\includegraphics[width=2.0\columnwidth]{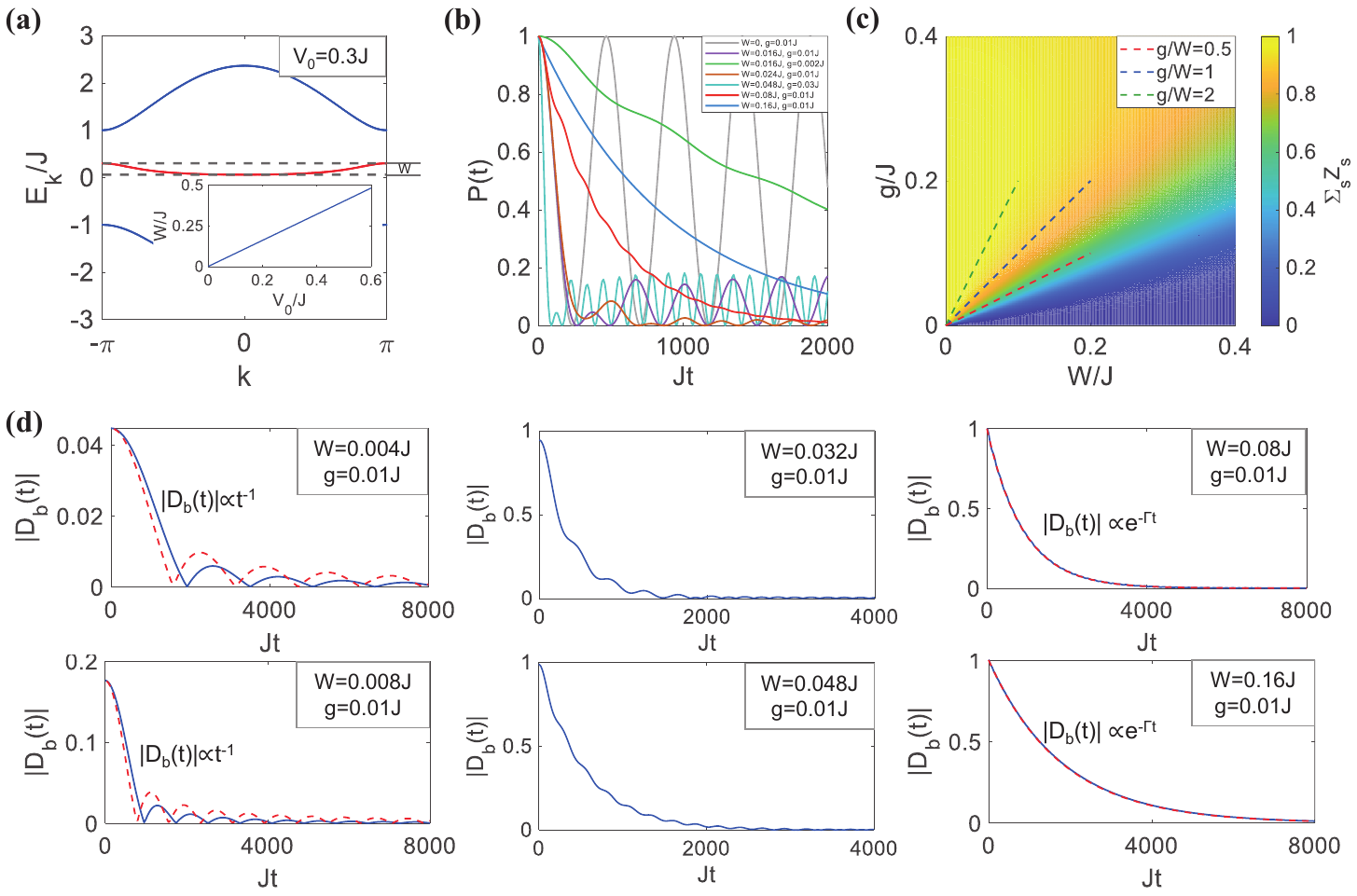}
		\caption{(Color online) Spontaneous emission dynamics in the modified 1D Lieb lattice with a narrow band. (a) Band structure of the modified model with $V_0=0.3J$. The formerly flat band becomes a narrow dispersive band (red) with bandwidth $W$. Inset: Bandwidth $W$ as a function of $V_0$, showing the linear relationship $W=\alpha V_0$ with $\alpha\approx0.802$. (b) Emitter population $P(t)$ for different parameters, showing the transition from non-Markovian oscillations to Markovian decay under different bandwidth and coupling strength. The emitter detuning $\Delta$ is fixed at the center of the narrow band in all cases. (c) Total bound-state residue $\sum_sZ_s$ as a function of $V_0$ for different coupling strengths $g$. Dashed lines indicate the scaling relation with $g/V_0$. (d) Branch-cut contribution $\left|D_b(t)\right|$ for $g=0.01J$ and different $V_0$, showing the transition from a power-law decay with oscillations ($W<g$) to an exponential decay ($g\ll W$) . Solid blue lines: numerical results. Dashed red lines: analytical approximations from Eqs.~(\ref{eq15}) and (\ref{eq16}).}
		\label{fig4}
	\end{figure*}

	In this section, we now turn to a more experimentally relevant scenario by introducing a controllable symmetry-breaking perturbation. This modification transforms the flat band into a narrow dispersive band, allowing us to systematically investigate how a finite bandwidth alters the spontaneous emission dynamics and governs the crossover from coherent trapping to dissipative decay. The Hamiltonian of this modified 1D Lieb lattice model is given by
	\begin{equation}
		H_b=\sum_j J\left(b_j^{A\dagger}b_j^B+b_j^{B\dagger}b_j^C+b_j^{B\dagger}b_{j+1}^C+h.c.\right)+V_0b_j^{C\dagger}b_j^C,
		\label{eq13}
	\end{equation}
	where $V_0$ is the on-site potential of the $C$ sublattice. This modification breaks the symmetry that protected the pristine flat band, transforming it into a narrow dispersive band. The energy band structure for $V_0=0.3J$ is shown in Fig.~\ref{fig4}(a). The formerly flat band (red) now exhibits a narrow but finite bandwidth, while the dispersive bands are also shifted in energy. The inset of Fig.~\ref{fig4}(a) quantifies the relationship between bandwidth of the narrow band and the on-site energy, that is $W=\alpha V_0$ with $\alpha\approx0.802$.
	
	The spontaneous emission dynamics of the emitter can be calculated by $P(t) = \left|\langle\psi(0)| e^{-iHt}|\psi(0)\rangle\right|^2$, numerically, with the initial state $|\psi(0)\rangle$ representing the excited emitter and the bath in vacuum. In all simulations within this section, the emitter's detuning $\Delta$ is set to the center of the emerging narrow band. This ensures that the observed dynamical transitions are purely due to the change in band dispersion, not a shift in emitter resonance. The numerical results, plotted in Fig.~\ref{fig4}(b), reveal a remarkable transition in the dynamics as the band from flat to dispersive. For the pristine flat band ($W=0$, gray line), the dynamics exhibits persistent, large-amplitude oscillations, a hallmark of coherent energy exchange with the localized flat-band states. Introducing a small finite bandwidth dramatically alters this behavior. For a fixed coupling $g=0.01J$, increasing bandwidth $W$ from $0.016J$ (purple line) to $0.08J$ (red line) monotonically reduces the long-time population and drives the system towards a Markovian decay regime. Intriguingly, a further increase of the bandwidth (blue line for $W=0.16J$) results in a significantly slower decay rate. Conversely, for a fixed nonzero bandwidth ($W=0.016J$), reducing the coupling strength to $g=0.002J$ (green line) also suppresses the population trapping, leading to a nearly complete Markovian decay. This nonmonotonic dependence of the decay rate on bandwidth is a key finding, indicating a complex competition between the coupling strength $g$ and the bandwidth of the narrow band $W$, which we analyze next.
	
	This complex behavior, particularly the nonmonotonic decay rate, can be understood by analyzing the Green's function of the system. For the emitter coupled to the $C$ sublattice of this modified lattice, the self-energy is given by (derived in Appendix~\ref{appendix:A})
	\begin{equation}
		\Sigma_s(\omega)=\frac{g^2(\omega^2-J^2)\theta(1-|z_-|)}{\sqrt{C^2-4\omega^2J^4}}-\frac{g^2(\omega^2-J^2)\theta(1-|z_+|)}{\sqrt{C^2-4\omega^2J^4}},
		\label{eq14}
	\end{equation}
	where $C=\omega^2(\omega-V_0)-J^2(\omega-V_0)-2J^2\omega$. The long-time population is governed by the residues $Z_s$ of the bound states that flank the narrow band. The evolution of the total residue $\sum_sZ_s$ under different $W$ and $g$ is shown in Fig.~\ref{fig4}(c). For a fixed bandwidth, the total residue increases with the coupling strength $g$, as a stronger $g$ promotes more robust bound states. For a fixed $g$, the total residue decreases monotonically as the bandwidth increases. These trends are fully consistent with the spontaneous emission dynamics in Fig.~\ref{fig4}(b). In the weak coupling limit ($g\ll W\ll J$), analytical analysis (derived in Appendix~\ref{appendix:D}) shows that $\sum_sZ_s$ becomes negligible. For instance, at $g=0.01J$ and $W=0.08J$, the total residue is only $\sum_sZ_s\approx0.0015$. This is a central finding: As the flat band is transformed into a narrow dispersive band, the system undergoes a systematic transition from a regime dominated by discrete, long-lived bound states (high residue, strong trapping) to one dominated by dissipation into the continuum (low residue, complete decay). Furthermore, the dashed lines in Fig.~\ref{fig4}(c) show that the total residue follows a universal scaling relation as a function of the ratio $g/W$. This scaling law provide quantitative criteria to distinguish between the two dynamical regimes: The band acts as "effectively flat" when $g/W\gg1$, leading to dominant bound-state contributions and coherent oscillations; whereas it becomes "effectively dispersive" when $g/W\ll1$, resulting in Markovian exponential decay into the continuum. The nonmonotonic decay rate observed at intermediate marks the crossover between these two regimes.
	
	We now turn to the initial transient dynamics, captured by the branch cut contribution $D_b(t)$ of the Green's function [second term in Eq.~(\ref{eq6})]. We calculate $D_b(t)$ numerically for a fixed $g=0.01J$ and different bandwidth $W$, as shown in Fig.~\ref{fig4}(d). Our analysis identifies two distinct dynamical regimes. When the bandwidth is smaller than the coupling strength ($W<g\ll J$), $D_b(t)$ exhibits damped oscillations shown in the left column of Fig.~\ref{fig4}(d). In this regime, the emitter's dynamics can be approximated by (derived in Appendix~\ref{appendix:D})
	\begin{equation}
		G_s(t)\approx\sum_sZ_se^{-iE_st}+\frac{2A_0}{Wt}\sin\left(\frac{W}{2}t\right),
		\label{eq15}
	\end{equation}
	where $A_0=\int_{w_1}^{w_2}\textrm{d}\omega A(\omega)$. The analytical results from the second term of Eq.~(\ref{eq15}) [red dashed lines, left column of Fig.~\ref{fig4}(d)] and the numerical results calculated by the second term of Eq.~(\ref{eq6}) [blue solid lines, left column of Fig.~\ref{fig4}(d)] follow similar trends (discussed in Appendix~\ref{appendix:D}). This describes a new kind of emitter's dynamics which is described by a power-law decay $|D_b(t)|\propto (Wt)^{-1}$ modulated by coherent oscillations with a frequency proportional to $W$, which contributes to the spontaneous emission dynamics. As $W$ increases within this regime, the oscillation frequency rises and the decay accelerates.
	
	When the coupling is weak compared to the bandwidth ($g\ll W\ll J$), $D_b(t)$ exhibits nearly pure exponential decay, characteristic of Markovian dynamics, as shown in the right column of Fig.~\ref{fig4}(d). In this regime, the spontaneous emission dynamics can be only governed by the second term of Eq.~(\ref{eq6}) because the total residue of the bound states is approximately zero. The decay rate of the spontaneous emission dynamics $\Gamma$ can be extracted from the pole of $G_s(\omega)$ on the second Riemann sheet. The analytical derivation yields (derivced in Appendix~\ref{appendix:D})
	\begin{equation}
		\Gamma\approx\textrm{Im}\left[\frac{g^2}{V_0\sqrt{5\beta^2-6\beta+1}}\right],
		\label{eq16}
	\end{equation}
	where $\beta=\Delta/V_0\approx0.6$. This reveals a key scaling: the decay rate is proportional to $g^2/V_0\propto g^2/W$. The spontaneous emission dynamics follows $P(t)\approx |e^{-\Gamma t}|^2$, which is the Markovian decay. The analytical results from Eq.~(\ref{eq16}) [dashed red lines in the right column of Fig.~\ref{fig4}(d)] show good agreement with the numerical results [blue solid lines, right column of Fig.~\ref{fig4}(d)].
	
	The transition of $D_b(t)$ from oscillatory to exponential decay, as $W$ increases at fixed $g$, is clearly observed in Fig.~\ref{fig4}(d). Furthermore, the decay rate exhibits a non-monotonic dependence on $W$. It first increases with $W$ because the nascent narrow band concentrates a significant spectral weight [as captured by the factor $A_0$ in Eq.~(\ref{eq15})], enhancing the effective emitter-bath coupling. However, after reaching a maximum, it subsequently decreases as the band widens further because the density of states spreads out and the average coupling strength at the emitter's frequency effectively decreases.
	
	Combining the insights from Figs.~\ref{fig4}(c) and (d), we demonstrate that the emitter’s dynamics, from long-lived bound-state trapping to Markovian decay, is precisely tunable by the competition between the coupling strength $g$ and the engineered bandwidth $W$. This parameter control unlocks distinct functional regimes: An intermediate choice can be leveraged to achieve fast spontaneous emission while maintaining a minimal long-time population, which is crucial for high-performance single-photon sources, while the flat-band limit enables strong population trapping. Crucially, our framework provides a unified physical picture that directly explains recent experiments, such as the accelerated exponential decay observed in moir$\rm\acute{e}$ photonic crystal as arising from the Markovian regime~\cite{wang2025moire}, and more broadly, bridges the idealized coherent physics of flat bands with the dissipative dynamics of realistic narrow-band systems. By establishing this continuous pathway and supplying quantitative scaling laws, our work transforms qualitative analogy into a predictive toolkit for interpreting and engineering quantum emission across structured photonic environments.
		
	\section{Conclusion}
	In summary, we established a comprehensive and predictive theoretical framework that unified the coherent dynamics of ideal flat bands with the dissipative behavior of realistic narrow-band systems. Beyond confirming the sublattice-dependent nature of coupling, our analysis revealed that coupling to the $A$ or $C$ sublattice while resonating with the flat band activated a collective, size-independent interaction with the entire band, a mechanism fundamentally distinct from dispersive-band-mediated bound-state formation. Furthermore, by controllably introducing lattice symmetry breaking, we demonstrate and quantitatively mapping a continuous transition from non-Markovian to Markovian emission dynamics. This crossover was governed by a universal competition between the intrinsic coupling strength $g$ and the engineered bandwidth $W$, encapsulated in explicit scaling laws that serve as a practical guide for distinguishing "effectively flat" from "effectively dispersive" bands. It provides a predictive toolkit that empowers the accurate interpretation of complex experiments in platforms like moir$\rm\acute{e}$ photonic crystals and, looking forward, offers a clear pathway for engineering desired quantum emission properties, from coherent photon trapping to high-speed single-photon generation, across a wide spectrum of structured photonic environments.
	
	\section*{Acknowledgements}
	This research is funded by the Key Project of the National Natural Science Foundation of China Joint Funds (No. U25A20197), the National Natural Science Foundation of China (No. 12374246). Y.H. acknowledges support by Beijing National Laboratory for Condensed Matter Physics (No. 2023BNLCMPKF001).
	
	\section*{Data avalilability}
	The data that support the findings of this article are openly available~\cite{data2026}.
	
	\appendix
	\renewcommand{\thesection}{\Alph{section}}
	\section{Derivation of the single particle Green's function of the systems}
	\label{appendix:A}
	In this appendix, we derive the single-particle Green's function $G_s(\omega)$. The spontaneous emission dynamics is charactrized by the temporal evolution of the emitter's excitd-state population, $P(t)=\langle a^\dagger(t)a(t)\rangle$ with $P(0)=1$, which can be written as
	\begin{equation}
		P(t)=|G_s(t)|^2,
		\label{eqA1}
	\end{equation}
	where $G_s(t)=-i\langle0|a(t)a^\dagger(0)|0\rangle$ for $t>0$ is the retarded Green’s function evaluated in the vacuum state. The bound state energies $E_s$ [determined $G_s^{-1}(E_s)=0$] and the spectrum function $A(\omega)=\frac{1}{\uppi}\textrm{Im}\left[G_s(\omega+i0^+)\right]$ follow from the frequency-domain Green’s function
	\begin{equation}
		G_s(\omega)=\frac{1}{\omega-\Delta-\Sigma_s(\omega)},
		\label{eqA2}
	\end{equation}
	where $\Delta$ is the detuning of the emitter and $\Sigma_s(\omega)$ is the self-energy, which can be expressed as
	\begin{equation}
		\Sigma_s(\omega)=\sum_{k}h_c^\dagger\frac{1}{\omega I-h_b(k)}h_c,
		\label{eqA3}
	\end{equation}
	where $I$ is the $3\times3$ identity matrix and $h_b$ is the bath Hamiltoinian matrix in momentum space with the basis $b_k=(b_k^A,b_k^B,b_k^C)^T$ which $b_k$ is the annihilation operator of the bath $k$ mode. For the pristine 1D Lieb lattice,
	\begin{equation}
		h_b(k)=\left(\begin{array}{ccc}
			{0} & {J} & {0} \\
			{J} & {0} & {J(1+e^{ik})} \\
			{0} & {J(1+e^{-ik})} & {0}
		\end{array}\right).
		\label{eqA4}
	\end{equation}
	The coupling vector $h_c$ encodes the emitter‑bath interaction with basis $b_k=(b_k^A,b_k^B,b_k^C)^T$. Its form depends on the sublattice to which the emitter is coupled.
	
	For the emitter couples to the $B$ sublattice of the 1D Lieb model, the Hamiltonian $h_c=(0,\frac{g}{\sqrt{N}},0)^T$. From Eq.~(\ref{eqA3}), the self-energy $\Sigma_s(\omega)$ is
	\begin{equation}
		\Sigma_s(\omega)=\frac{g^2\omega\theta(1-|z_-|)}{\sqrt{(\omega^2-3J^2)^2-4J^4}}-\frac{g^2\omega\theta(1-|z_+|)}{\sqrt{(\omega^2-3J^2)^2-4J^4}},
		\label{eqA5}
	\end{equation}
	where $\theta(1-|z_\pm|)$ is the step function with $z_\pm=\frac{\omega^2-3J^2\pm\sqrt{(\omega^2-3J^2)^2-4J^4}}{2J^2}$. For the emitter couples to the $C$ sublattice of the 1D Lieb model, the Hamiltonian $h_c=(0,0,\frac{g}{\sqrt{N}})^T$. From Eq.~(\ref{eqA3}), the self-energy $\Sigma_s(\omega)$ is
	\begin{equation}
		\Sigma_s(\omega)=\frac{g^2(\omega^2-J^2)\theta(1-|z_-|)}{\omega\sqrt{(\omega^2-3J^2)^2-4J^4}}-\frac{g^2(\omega^2-J^2)\theta(1-|z_+|)}{\omega\sqrt{(\omega^2-3J^2)^2-4J^4}}.
		\label{eqA6}
	\end{equation}
	For the emitter couples to the $A$ sublattice of the 1D Lieb model, the Hamiltonian $h_c=(\frac{g}{\sqrt{N}},0,0)^T$. From Eq.~(\ref{eqA3}), the self-energy $\Sigma_s(\omega)$ is
	\begin{equation}
		\begin{split}
			\Sigma_s(\omega)=&\frac{g^2(\omega^2-2J^2(1+z_-))\theta(1-|z_-|)}{\omega\sqrt{(\omega^2-3J^2)^2-4J^4}} \\
			&-\frac{g^2(\omega^2-2J^2(1+z_+))\theta(1-|z_+|)}{\omega\sqrt{(\omega^2-3J^2)^2-4J^4}},
		\end{split}
		\label{eqA7}
	\end{equation}
	which we discuss in Appendix~\ref{appendix:C}.
	
	When an on‑site potential $V_0$ is added on the $C$ sublattice, the bath matrix becomes
	\begin{equation}
		\begin{split}
			h_b(k)=\left(\begin{array}{ccc}
				{0} & {J} & {0} \\
				{J} & {0} & {J(1+e^{ik})} \\
				{0} & {J(1+e^{-ik})} & {V_0}
			\end{array}\right).
		\end{split}
		\label{eqA8}
	\end{equation}
	Taking again $h_c=(0,0,\frac{g}{\sqrt{N}})^T$, the self‑energy reads
	\begin{equation}
		\Sigma_s(\omega)=\frac{g^2(\omega^2-J^2)\theta(1-|z_-|)}{\sqrt{C^2-4\omega^2J^4}}-\frac{g^2(\omega^2-J^2)\theta(1-|z_+|)}{\sqrt{C^2-4\omega^2J^4}},
		\label{eqA9}
	\end{equation}
	where $C=\omega^2(\omega-V_0)-J^2(\omega-V_0)-2J^2\omega$. Inserting the appropriate self‑energy Eq.~(\ref{eqA5}) to (\ref{eqA7}) and (\ref{eqA9}) into Eq.~(\ref{eqA2}) yields the Green’s function for each configuration, from which the corresponding spontaneous‑emission dynamics can be calculated.
	
	\renewcommand{\thesection}{\Alph{section}}
	\section{Hamiltonian in band representation}
	\label{appendix:B}
	In this appendix, we derive the Hamiltonian in diagonalized band representation. The Hamiltonian Eq.~(\ref{eqA4}) can be written in the diagonal basis by a unitary transformation $U_k^\dagger h_b(k)U_k=\textrm{diag}(E_k^{(+)},E_k^{(0)},E_k^{(-)})$, which yields the band energies
	\begin{equation}
		E_k^{(\pm)}=\pm\sqrt{3+2\cos(k)},\ E_k^{(0)}=0.
		\label{eqB1}
	\end{equation}
	The unitary matrix $U_k=(u_k^{(+)},u_k^{(0)},u_k^{(-)})$ is composed of the normalized eigenvectors
	\begin{equation}
		\begin{split}
			u_k^{(+)}&=\sqrt{\frac{1+\cos(k)}{3+2\cos(k)}}\left(\begin{array}{c}
				{\frac{1}{1+e^{-ik}}} \\
				{\frac{\sqrt{3+2\cos(k)}}{1+e^{-ik}}} \\
				{1}
			\end{array}\right), \\
			u_k^{(0)}&=\frac{1}{\sqrt{3+2\cos(k)}}\left(\begin{array}{c}
				{-1-e^{ik}} \\
				{0} \\
				{1}
			\end{array}\right), \\
			u_k^{(-)}&=\sqrt{\frac{1+\cos(k)}{3+2\cos(k)}}\left(\begin{array}{c}
				{\frac{1}{1+e^{-ik}}} \\
				{\frac{\sqrt{3+2\cos(k)}}{1+e^{-ik}}} \\
				{1}
			\end{array}\right).
		\end{split}
		\label{eqB2}
	\end{equation}
	The transformation to the band annihilation operators $\psi_k^{(n)}$ is defined by $\psi_k=U_k^*b_k$. Substituting this into $H_b=\sum_kb_k^\dagger h_b(k)b_k$ gives the diagonal form
	\begin{equation}
		H_b=\sum_{k,n}E_k^{(n)}\psi_k^{(n)\dagger}\psi_k^{(n)}.
		\label{eqB3}
	\end{equation}
	For an emitter coupled to sublattice $\gamma$ at unit cell $j=0$ with interaction $H_c=g(a^\dagger b_0^\gamma+\rm{H.c.})$, we express $b_0^\gamma=\frac{1}{\sqrt{N}}\sum_kb_k^\gamma$ and transform to the band basis, obtaining
	\begin{equation}
		H_c=\sum_{k,n}g_k^{(n)}(a^\dagger\psi_k^{(n)}+\rm{H.c.}),
		\label{eqB4}
	\end{equation}
	where $g_k^{(n)}=\frac{g}{\sqrt{N}}U_{k,\gamma n}^*$ is the coupling strength in the band representation.
	
	For the emitter coupled to the $B$ sublattice, the coupling strength $g_k^{(n)}$ are
	\begin{equation}
		g_k^{(\pm)}=\frac{g}{\sqrt{N}}\cdot\frac{\sqrt{1+\cos(k)}}{1+e^{-ik}},\ g_k^{(0)}=0.
		\label{eqB5}
	\end{equation}
	The vanishing coupling to the flat band ($g_k^{(0)}=0$) reflects the dark-state nature of the $B$ sublattice.
	
	For the emitter coulpled to the $C$ sublattice, the coupling strength $g_k^{(n)}$ are
	\begin{equation}
		g_k^{(\pm)}=\frac{g}{\sqrt{N}}\sqrt{\frac{1+\cos(k)}{3+2\cos(k)}},\  g_k^{(0)}=\frac{g}{\sqrt{N}}\frac{1}{\sqrt{3+2\cos(k)}}.
		\label{eqB6}
	\end{equation}
	In this case, the interaction with the flat band can be written as
	\begin{equation}
		H_c=\frac{g}{\sqrt{N}}\sum_k\frac{1}{\sqrt{3+2\cos(k)}}(a^\dagger\psi_k^{(0)}+\mathrm{H.c.}).
		\label{eqB7}
	\end{equation}
	When the emitter resonates with the flat band, this interaction can be recast in terms of a collective mode $B=\frac{\sum_k g_k^{(0)}}{\sqrt{\sum_k|g_k^{(0)}|^2}}\psi_k^{(0)}$, leading to
	\begin{equation}
		H_c=g_B(a^\dagger B+B^\dagger a),
		\label{eqB8}
	\end{equation}
	where the coupling strength of the emitter and the collective modes of the flat band is $g_B=\sqrt{\sum_k|g_k^{(0)}|^2}=\frac{g}{5^{1/4}}$, which is independent of the system size $N$.
	
	For the emitter coupled to the $A$ sublattice, the coupling strength $g_k^{(n)}$ are
	\begin{equation}
		\begin{split}
			g_k^{(\pm)}&=\frac{g}{\sqrt{N}(1+e^{-ik})}\sqrt{\frac{1+\cos(k)}{3+2\cos(k)}} \\
			g_k^{(0)}&=\frac{g}{\sqrt{N}}\frac{1+e^{ik}}{\sqrt{3+2\cos(k)}}.
		\end{split}
		\label{eqB9}
	\end{equation}
	Similarly, when the emitter resonates with the flat band, the interaction can be expressed via a collective mode with an effective coupling strength $g_B=g\sqrt{1-\frac{1}{\sqrt{5}}}$, which is also size independent.
	
	\renewcommand{\thesection}{\Alph{section}}
	\section{Spontaneous emission of the emitter couples to A sublattice}
	\label{appendix:C}
	
	\begin{figure*}[t]
		\includegraphics[width=2.0\columnwidth]{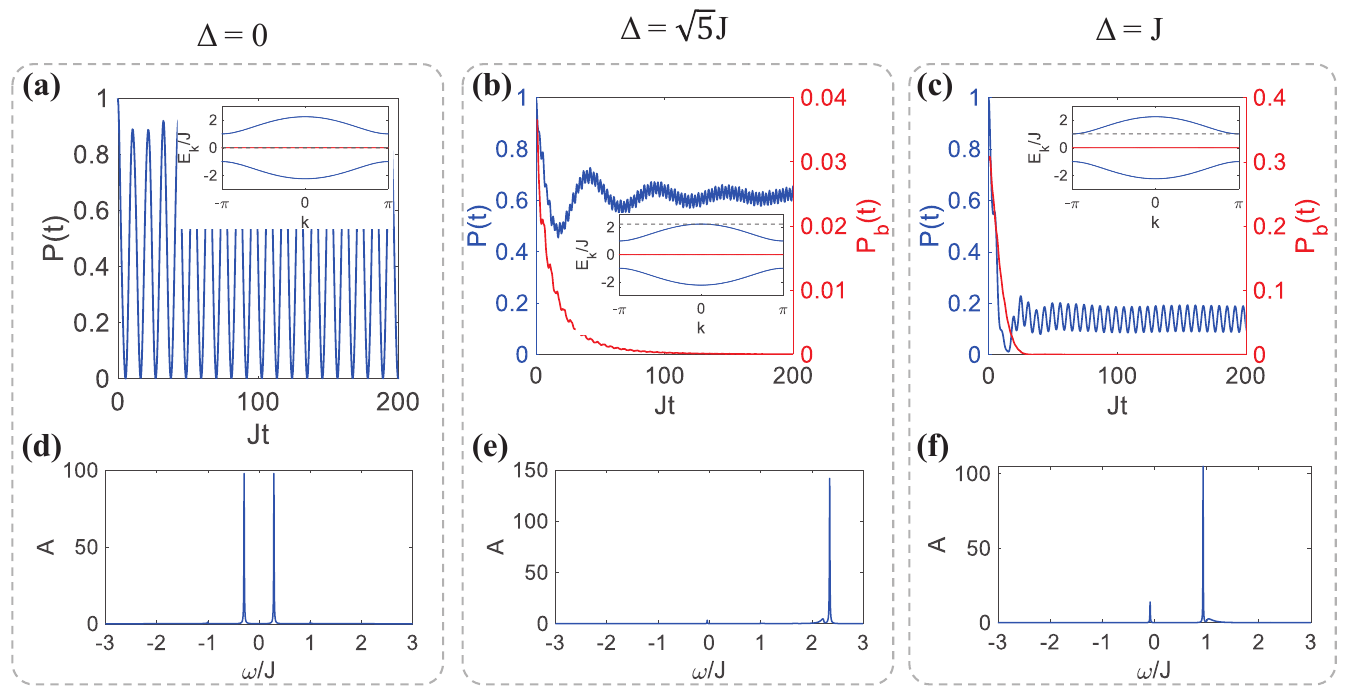}
		\caption{(Color online) Dynamics and spectral properties for an emitter coupled to the $A$ sublattice with $g=0.4J$. (a)-(c) Spontaneous emission dynamics $P(t)$ (blue lines) for (a) $\Delta=0$, (b) $\Delta=\sqrt{5}J$ and (c) $\Delta=J$. The red lines represent the continuum contribution $P_b(t)=\left|D_b(t)\right|^2$. Insets: Schematic of the emitter detuning (gray dashed line) relative to the bath bands. (d)-(f) The corresponding spectral function $A(\omega)$ for each detuning case in (a)-(c). Sharp peaks correspond to bound states outside the continuum, while broad features originate from the dispersive band continuum.}
		\label{figC1}
	\end{figure*}
	
	In this appendix, we consider the case that the single emitter couples to the sublattice $A$ in 1D Lieb model. The interaction Hamiltonian is
	\begin{equation}
		H_c=g\left(a^\dagger b_0^A+b_0^{A\dagger}a\right),
		\label{eqC1}
	\end{equation}
	where $g$ is the emitter-bath coupling strength. For the emitter coupled to the $A$ sublattice, the self-energy is found to be
	\begin{equation}
		\begin{split}
			\Sigma_s(\omega)=&\frac{g^2(\omega^2-2J^2(1+z_-))\theta(1-|z_-|)}{\omega\sqrt{(\omega^2-3J^2)^2-4J^4}} \\
			&-\frac{g^2(\omega^2-2J^2(1+z_+))\theta(1-|z_+|)}{\omega\sqrt{(\omega^2-3J^2)^2-4J^4}},
		\end{split}
		\label{eqC2}
	\end{equation}
	where $\theta(1-|z_\pm|)$ is the step function with $z_\pm=\frac{\omega^2-3J^2\pm\sqrt{(\omega^2-3J^2)^2-4J^4}}{2J^2}$. The bound states $E_s$ are determined by $G_s^{-1}(E_s)=0$.

	For $\Delta=0$, the emitter resonates with the flat band, and two in-gap bound states emerge at energies $E_\pm\approx\pm0.29J$. The total residue of these bound states is $\sum_{s=\pm}Z_s\approx0.95$, indicating that the spontaneous emission dynamics is dominated by their contribution, $P(t)=\left|\sum_{s=\pm}Z_se^{-iE_s t}\right|^2=\left|2Z_s\sin\left(\left|E_s\right|t\right)\right|^2$. The coherent oscillations evident in the numerical results of Fig.~\ref{figC1}(a) originate from the these two bound states. The oscillation period, $T=\frac{\uppi}{J\left|E_s\right|}\approx10.8/J$, determined by the bound-state energy, shows excellent agreement with the numerical simulation. Furthermore, the constant amplitude of these oscillations over a long time is consistent with the large total residue, signifying minimal population decay into the continuum. The spectral function $A(\omega)$, plotted in Fig.~\ref{figC1}(d), shows sharp peaks at the bound-state energies, while the flat band contributes negligible spectral weight, confirming its role in mediating the formation of bound states rather than irreversible decay.
	
	The formation mechanism of these bound states is fundamentally distinct from the $B$ sublattice case. Here, the flat band does not act as a passive reservoir but as a single, collective degree of freedom. In band representation, the flat-band Hamiltonian becomes Eq.~(\ref{eq9}) with $E_k=0$. The interaction Hamiltonian in the diagonal basis is Eq.~(\ref{eq10}), where $g_k$ is the coupling constant between the emitter and the bath mode $k$. For the flat band, the coupling strength is $g_k^{(\textrm{flat})}=\frac{g}{\sqrt{N}}\times\frac{1+e^{ik}}{\sqrt{3+2\cos(k)}}$, which can be collectively enhanced. In collective representation, the interaction Hamiltonian can be rewritten as $H_c=g_B\left(a^\dagger B+B^\dagger a\right)$, where $B=\frac{\sum_k g_k^{(\textrm{flat})}\psi_k}{\sqrt{\sum_k|g_k^{(\textrm{flat})}|^2}}$ is a collective mode encompassing the entire flat band, and the effective coupling strength $g_B=g\sqrt{1-\frac{1}{\sqrt{5}}}$ is independent of system size. This collective enhancement leads to the steady coherent oscillations in dynamics [shown in Fig.~\ref{figC1}(a)]. Therefore, at $\Delta=0$, the spontaneous emission dynamics differs markedly depending on the coupling sublattice: it is dominated by bound states arising from the dispersive bands when the emitter couples to the $B$ sublattice, in contrast to the case of $A$ sublattice coupling, where the dynamics is governed by bound states formed via the flat band.
	
	We further explore the detuning dependence of the emitter's dynamics. While the DOS diverges at both the dispersive band edges and the flat band, the spontaneous emission dynamics exhibits markedly distinct characteristics. For $\Delta=\sqrt{5}J$, the emitter resonates with the upper edge of the dispersive band [inset of Fig.~\ref{figC1}(b)]. The dynamics calculated numerically, shown by the blue line in Fig.~\ref{figC1}(b), is consistent with the analytical result from Eq.~(\ref{eq6}). In this case, one bound state resides inside the gap ($E_-\approx-0.039J$) and another lies above the continuum ($E_+\approx2.35J$). The long-time dynamics is dominated by the two bound states, $\left|\sum_{s=\pm}Z_se^{-iE_s t}\right|^2$. The oscillation period is $T=\frac{2\uppi}{J\left|E_+-E_-\right|}\approx2.39/J$ and the amplitude of these oscillations is governed by the residue $Z_+\approx0.79$ of bound state at $E_+\approx2.35J$, closer to the upper band edge, which is consistent with the numerical results [Fig.~\ref{figC1}(b)].
	
	An analogous discussion applies to the case of $\Delta=J$ [inset of Fig.~\ref{figC1}(c)], where one bound state lies below the flat band ($E_-\approx-0.082J$) and another above the upper dispersive band ($E_+\approx0.93J$). The dynamics calculated numerically, shown by the blue line in Fig.~\ref{figC1}(c), is consistent with the analytical result from Eq.~(\ref{eq6}). The long-time dynamics is dominated by the two bound states, $\left|\sum_{s=\pm}|Z_se^{-iE_s t}\right|^2$. The oscillation period is $T=\frac{2\uppi}{J\left|E_+-E_-\right|}\approx6.3/J$ and the amplitude of these oscillations is governed by the residue $Z_+\approx0.39$ of the bound state at $E_+\approx0.93J$, closer to the lower band edge, which are consistent with the numerical results [Fig.~\ref{figC1}(c)].
	
	The red lines in Figs.~\ref{figC1}(b) and \ref{figC1}(c) represent the continuum contribution $P_b(t)=\left|D_b(t)\right|^2$ by the second term of Eq.~(\ref{eq6}), which captures the spontaneous emission dynamics evolving from the initial condition. The decay rate of $P_b(t)$ is significantly smaller for $\Delta=\sqrt{5}J$ than for $\Delta=J$. This difference can be understood by examining the momentum dependence of the coupling $g_k$ given in Eq.~(\ref{eq10}) together with the computed spectral features. For the dispersive band, the coupling strength is $g_k^{(\textrm{disp})}=\frac{g}{\sqrt{N}(1+e^{-ik})}\times\sqrt{\frac{1+\cos(k)}{3+2\cos(k)}}$, which is highly asymmetric and increases monotonically with $k$ from 0 to $\uppi$. The spectral function $A(\omega)$ reveals that for $\Delta=\sqrt{5}J$ [shown in Fig.~\ref{figC1}(e)], two sharp peaks at the bound-state energies coexist with a broad continuum near the upper band edge of the dispersive band ($k\approx0$), where the coupling strength $g_k^{(\textrm{disp})}$ is relatively small. In contrast, for $\Delta=J$ [shown in Fig.~\ref{figC1}(f)], the spectral weight is concentrated near the lower band edge ($k\approx\uppi$), where the coupling strength $g_k$ is large. Therefore, the decay rate of $P_b(t)$ is significantly smaller for $\Delta=\sqrt{5}J$ than for $\Delta=J$, which is different from the case that the emitter couples to the C sublattice disscussed in the main text.
	
	The rich detuning dependence revealed in Figs.~\ref{figC1}(a) to \ref{figC1}(c) finds a unified explanation in the contrasting nature of emitter-band coupling. As in the $C$-sublattice case analyzed in the main text, coupling to the flat band occurs via a single collective mode with size-independent strength $g_B$, while coupling to the dispersive bands involves a sum over extended Bloch modes. Despite the divergent DOS at both the flat band and the dispersive band edges, their vastly different coupling strengths lead to markedly different dynamical regimes. The detuning $\Delta$ thus serves as a continuous parameter that reweights the emitter’s interaction between the robust collective flat-band channel and the dissipative dispersive continua, offering a tunable pathway between coherent trapping and Markovian decay.
	
	\begin{widetext}
		\renewcommand{\thesection}{\Alph{section}}
		\section{Derivation of the spontaneous emission dynamics}
		\label{appendix:D}
		
		In this appendix, we derive the spontaneous emission dynamics of the emitter analytically in two cases $W<g\ll J$ and $g\ll W\ll J$. The single-particle Green's function in Lehmann spectral representation is
		\begin{equation}
			G_s(t)=D_s(t)+D_b(t)=\sum_sZ_se^{-iE_st}+\int_{w_1}^{w_2}\textrm{d}\omega A(\omega)e^{-i\omega t},
			\label{eqD1}
		\end{equation}
		where $Z_s$ is the residue of the bound state $E_s$ and $A(\omega)$ is the energy spectrum of the Green's function. The integration limits $w_1$ and $w_2$ represent the band-bottom and band-top frequencies of the narrow band, respectively. The second term $D_b(t)$ of Eq.~(\ref{eqD1}) captures the initial transient of the spontaneous emission dynamics. $D_b(t)$ can be calculated by considering the expansion of $A(\omega)$ around the average value of $A(\omega)$
		\begin{equation}
			D_b(t)=A_0\int_{w_1}^{w_2}\textrm{d}\omega e^{-i\omega t}+\int_{w_1}^{w_2}\textrm{d}\omega\delta A(\omega)e^{-i\omega t},
			\label{eqD2}
		\end{equation}
		where $A_0=\frac{1}{W}\int_{w_1}^{w_2}\textrm{d}\omega A(\omega)$ with bandwidth $W$ is the average value of $A(\omega)$ and $\delta A(\omega)=A(\omega)-A_0$ is the fluctuation of the spectral function. $w_1$ and $w_2$ are the frequencies at the lower and upper band edges of the narrow band. The key to simplifying this expression lies in the second term, $R(t) = \int_{w_1}^{w_2}\textrm{d}\omega\delta A(\omega) e^{-i\omega t}$. As shown in Figs.~\ref{figD1}(a) and \ref{figD1}(b), $\delta A(\omega)$ is not random noise but a smooth function that deviates positively from zero in parts of the band and negatively in others. When the bandwidth $W$ is small, the phase factor $e^{-i\omega t}$ varies slowly across the interval $[w_1,w_2]$. It can be approximated by a constant (e.g., its value at the band center) over the entire integration range. Therefore,
		\begin{equation}
			R(t)=\int_{w_1}^{w_2}\textrm{d}\omega\delta A(\omega) e^{-i\omega t}\approx e^{-i\omega_ct}\int_{w_1}^{w_2}\textrm{d}\omega\delta A(\omega)=0,
			\label{eqD3}
		\end{equation}
		where $\omega_c$ is the central frequency of the band. This cancellation leads to the approximation
		\begin{equation}
			D_b(t)\approx A_0\int_{w_1}^{w_2}\textrm{d}\omega e^{-i\omega t}=\frac{2A_0}{t}\sin\left(\frac{W}{2}t\right).
			\label{eqD4}
		\end{equation}
		This analytical result captures the characteristic power-law decay $|D_{b}(t)| \propto t^{-1}$ modulated by coherent oscillations with frequency $W/2$, which originates from the finite bandwidth and the assumed sharp spectral cutoffs at $w_1$ and $w_2$. It is important to note that in a real physical system, the spectral function $A(\omega)$ vanishes smoothly at the band edges, as shown in Figs.~\ref{figD1}(a) and \ref{figD1}(b), rather than abruptly and $R(t)\neq0$. This smooth roll-off means the actual spectral weight lacks the high-frequency components present in our idealized rectangular model (which assumes a constant $A_0$ across the entire band). Consequently, the analytical approximation in Eq.~(\ref{eqD4}), which incorporates these unphysical high-frequency components, tends to overestimate the initial decay rate compared to the exact numerical dynamics. Nevertheless, Eq.~(\ref{eqD4}) provides a qualitatively correct and sufficient description of the dominant $D_{b}(t)$ dynamics, capturing both the power-law decay and the oscillation frequency.
		
		\begin{figure}[t]
			\includegraphics[width=0.8\columnwidth]{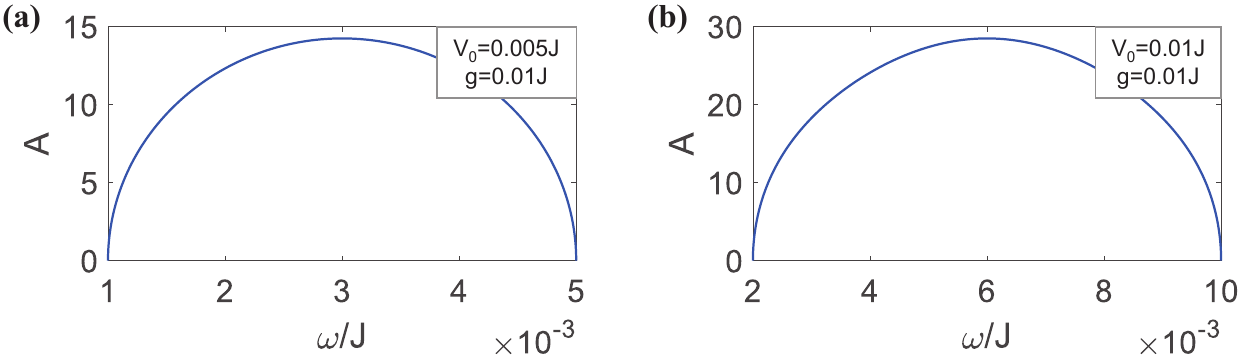}
			\caption{(Color online) The spectral function of the narrow band under (a) $V_0=0.005J$, $g=0.01J$; (b) $V_0=0.01J$, $g=0.01J$.}
			\label{figD1}
		\end{figure}
		
		Then we derive the residues of the bound states and the spontaneous emission decay rate for $g\ll V_0\ll J$. The single particle Green's function can be derived by path-integral method, which is
		\begin{equation}
			G_s(\omega)=\frac{1}{\omega-\Delta-\Sigma_s(\omega)},
			\label{eqD5}
		\end{equation}
		where the self-energy $\Sigma_s(\omega)$ is
		\begin{equation}
			\begin{split}
				\Sigma_s(\omega)&=g^2\int\frac{\textrm{d}k}{2\uppi}\frac{\omega^2-J^2}{C-\omega J^2 e^{ik}-\omega J^2e^{-ik}} \\
				&=\frac{g^2(\omega^2-J^2)\theta(1-|z_-|)}{\sqrt{C^2-4\omega^2J^4}}-\frac{g^2(\omega^2-J^2)\theta(1-|z_+|)}{\sqrt{C^2-4\omega^2J^4}},
			\end{split}
			\label{eqD6}
		\end{equation}
		where $C=\omega^2(\omega-V_0)-J^2(\omega-V_0)-2J^2\omega$. From the numerical results, the poles of the single particle Green's function are $E_+=V_0+\delta$ and $E_-=V_0-W-\delta$, where $W$ is the bandwidth of the narrow band and $\delta$ is a small shift with $\delta>0$ and $\delta\ll V_0$. For the case $E_+=V_0+\delta$, the derivative of the self-energy $\partial_\omega\Sigma_s(\omega)$ is
		\begin{equation}
			\begin{split}
				\partial_\omega\Sigma_s(\omega)&=\frac{g^2(\omega^2-J^2)(2(\omega^2+2\omega(\omega-V_0)-3J^2)-8J^4\omega)(\omega^2(\omega-V_0)-J^2(\omega-V_0)-2J^2\omega)}{C^{3/2}}-\frac{2g^2\omega}{\sqrt{C}} \\
				&\approx -\frac{g^2V_0}{(V_0\delta)^{3/2}}-\frac{g^2V_0}{J^2\sqrt{V_0\delta}}.
			\end{split}
			\label{eqD7}
		\end{equation}
		From the above equation, the derivative of the self-energy $\partial_\omega\Sigma_s(\omega)$ tends to infinity and the residue of this pole $Z_+=(1-\partial_\omega\Sigma_s(\omega)|_{\omega=E_+})^{-1}\approx0$. For the case $E_-=V_0-W-\delta$, the derivative of the self-energy $\partial_\omega\Sigma_s(\omega)$ is
		\begin{equation}
			\begin{split}
				\partial_\omega\Sigma_s(\omega)&=-\frac{g^2(\omega^2-J^2)(2(\omega^2+2\omega(\omega-V_0)-3J^2)-8J^4\omega)(\omega^2(\omega-V_0)-J^2(\omega-V_0)-2J^2\omega)}{C^{3/2}}+\frac{2g^2\omega}{\sqrt{C}} \\
				&\approx -\frac{3g^2J^6(3W-2V_0)}{(5J^4W^2-4J^4V_0W)^{3/2}}+\frac{2g^2(V_0-W)}{\sqrt{5J^4W^2-4J^4V_0W}} \\
				&=-\frac{3g^2(3\alpha-2)}{V_0^2(5\alpha^2-4\alpha)^{3/2}}+\frac{2g^2(1-\alpha)}{J^2\sqrt{5\alpha^2-4\alpha}},
			\end{split}
			\label{eqD8}
		\end{equation}
		where $W=\alpha V_0$ with $\alpha\approx0.802$ as shown in Fig.~\ref{fig4}(a) in the main text. From Eq.~(\ref{eqD8}), the derivative of the self-energy $\partial_\omega\Sigma_s(\omega)$ tends to infinity and the residue of this pole $Z_-=(1-\partial_\omega\Sigma_s(\omega)|_{\omega=E_-})^{-1}\approx0$. Combining  Eq.~(\ref{eqD7}) and (\ref{eqD8}), the contribution of bound states to the emitter's population is almost negligible.
		
		The decay rate $\Gamma$ of the emitter's population can be extracted from the pole of the self-energy on the second Riemann sheet. From the numerical results, this pole of the single particle Green's function is $E_s=\Delta+\delta$ where $\delta\ll\Delta$ and $\delta$ is complex. The self-energy $\Sigma_s(\omega)$ becomes
		\begin{equation}
			\begin{split}
				\Sigma_s(\omega)&=-\frac{g^2(\omega^2-J^2)}{\sqrt{C^2-4\omega^2J^4}} \\
				&\approx \frac{g^2(2\beta V_0\delta-J^2)}{J^2\sqrt{(5\beta^2-6\beta+1)V_0^2+(10\beta-6)V_0\delta}},
			\end{split}
			\label{eqD9}
		\end{equation}
		where $\beta=\Delta/V_0\approx0.6$. The pole of the Green's function on the second Riemann sheet can be calculated by $G_s^{-1}(E_s)=0$ and we have
		\begin{equation}
			\begin{split}
				&\delta-\frac{g^2(2\beta V_0\delta-J^2)}{J^2\sqrt{(5\beta^2-6\beta+1)V_0^2+(10\beta-6)V_0\delta}}=0, \\
				&\delta\approx \frac{g^2}{V_0\sqrt{5\beta^2-6\beta+1}}.
			\end{split}
			\label{eqD10}
		\end{equation}
		The decay rate of the population is $\Gamma=\textrm{Im}\left[\frac{g^2}{V_0\sqrt{5\beta^2-6\beta+1}}\right]$ and the spontaneous emission dynamics of the emitter is $P(t)\approx|e^{-\Gamma t}|^2$ under $g\ll V_0\ll J$.
	\end{widetext}


\begin{thebibliography}{48}
\bibitem [{dat()}]{data2026}%
		\BibitemOpen
		\href@noop {} {\bibinfo {title} {The computational codes that support the
				findings of this study are available upon reasonable request.}}\BibitemShut
		{Stop}%
\end{thebibliography}
\end{document}